\documentclass[aps,prd,reprint,nofootinbib,groupedaddress,preprintnumbers,longbibliography]{revtex4-2}

\usepackage{amsmath,amssymb,mathtools,bm}
\usepackage{graphicx,color}
\usepackage[dvipsnames]{xcolor}
\usepackage{float}
\usepackage{tensor}
\usepackage{ulem}

\usepackage{hyperref}
\hypersetup{
    colorlinks=true,
    linkcolor=blue,
    citecolor=blue,
    filecolor=magenta,
    urlcolor=blue
}

\begin{document}

\title{Ultralight Dark Matter Statistics for Pulsar Timing Detection}

\author{Kimberly K.~Boddy}
\email{kboddy@physics.utexas.edu}
\affiliation{Texas Center for Cosmology and Astroparticle Physics, Weinberg Institute, Department of Physics, The University of Texas at Austin, Austin, TX 78712, USA}

\author{Jeff A.~Dror} 
\email{jeffdror@ufl.edu}
\affiliation{Institute for Fundamental Theory, Physics Department, University of Florida, Gainesville, FL 32611, USA}

\author{Austin Lam}
\email{a.lam2@ufl.edu}
\affiliation{Institute for Fundamental Theory, Physics Department, University of Florida, Gainesville, FL 32611, USA}

\begin{abstract}
Fluctuations in ultralight dark matter produce significant metric perturbations, which may be detected by monitoring the arrival times of light from millisecond pulsars. While searches using this technique are already underway, they do not consistently account for the statistical properties of the dark matter field. The statistics of this field depend on the velocity dispersion of dark matter and, consequently, its coherence length. In the mass range relevant for pulsar timing arrays, the coherence length is comparable to separations between pulsars, making it crucial to incorporate its effects into the analysis. This work presents a consistent statistical method for gravitational direct detection of ultralight dark matter. Our key result is the derivation of the two-point function of the metric fluctuations, which we apply to pulsar timing and discuss its implementation in future searches.
\end{abstract}

\maketitle

\vspace{0.2cm}
\noindent \textbf{Introduction.}
While the fundamental nature of dark matter remains unknown, a well-motivated possibility is that it comprises a light bosonic field~\cite{Antypas:2022asj}. If the particles associated with this field have a de Broglie wavelength on the scale of galaxies, they generate significant perturbations in the spacetime metric. These perturbations are analogous to gravitational waves---in the sense that they are oscillating weak-field corrections to the Minkowski metric---and have a frequency approximately equal to twice the dark matter mass. These oscillations present an opportunity for {\it gravitational direct detection} of dark matter, leveraging the existing infrastructure for gravitational wave detection.

The potential significance of these metric perturbations was first highlighted by Khmelnitsky and Rubakov~\cite{Khmelnitsky:2013lxt}, sparking considerable theoretical interest in their detection through methods such as pulsar timing~\cite{Khmelnitsky:2013lxt,Porayko:2014rfa, Graham:2015ifn, Aoki:2016mtn,DeMartino:2017qsa,Kato:2019bqz,Nomura:2019cvc,Kaplan:2022lmz,Unal:2022ooa,Xia:2023hov,Luu:2023rgg,Hwang:2023odi,Kim:2023kyy}, laser interferometry~\cite{Aoki:2016kwl,Kim:2023pkx,Brax:2024yqh,Yu:2024enm}, astrometry~\cite{Dror:2024con,Kim:2024xcr}, and their influence on gravitational-wave sources~\cite{Blas:2016ddr,Blas:2024duy}.

Pulsar timing array (PTA) experiments, which primarily aim to detect low-frequency gravitational waves, are sensitive to ultralight dark matter perturbations for dark matter masses in the range of $m \sim 10^{-24}~\mathrm{eV} $ -- $ 10^{-22}~\mathrm{eV} $. The PPTA~\cite{Porayko:2018sfa}, EPTA~\cite{EPTA:2023xxk,EPTA:2023xiy}, and NANOGrav~\cite{NANOGrav:2023hvm} collaborations have performed searches for ultralight dark matter and are already sensitive to a dark matter density comparable to its expected local value of $\sim 0.3~\mathrm{GeV/cm^3}$~\cite{2017PASJ...69R...1S}. While small-scale structure observations constrain ultralight dark matter masses to be above $ 10^{-22}$ -- $10^{-20}~\mathrm{eV}$~\cite{Kobayashi:2017jcf, Irsic:2017yje, Nori:2018pka, Leong:2018opi, Schutz:2020jox, DES:2020fxi, Rogers:2020ltq,Dalal:2022rmp}, a subcomponent of dark matter could be in the form of ultralight particles with a smaller mass.

Searches for these metric perturbations by PTAs hold great promise for advancing our understanding of dark matter. Qualitatively, the effectiveness of pulsar timing searches for dark matter is governed by the comparison of the dark matter coherence length to the distance between the Earth and pulsars or between different pulsars.
The coherence length is given by~\cite{Khmelnitsky:2013lxt}
\begin{equation}
  \ell \equiv \frac{1}{m v_0} \simeq 0.87~\mathrm{kpc} \left( \frac{10^{-23}~\mathrm{eV}}{m} \right) \left( \frac{220~\mathrm{km/s}}{v_0} \right) \, ,
  \label{eq:coherence_length}
\end{equation}
where $v_0$ is the local velocity dispersion of dark matter. A common assumption is that dark matter-induced metric perturbations are either uncorrelated or fully correlated at each observation point (Earth or pulsar)~\cite{Porayko:2018sfa,EPTA:2023xxk,EPTA:2023xiy,NANOGrav:2023hvm}. The approximation of uncorrelated perturbations holds when the observer-pulsar separations significantly exceed the coherence length, and vice versa for fully correlated perturbations; however, these approximations are rarely true for an entire set of PTA pulsars, where typical separations range from $100~\mathrm{pc}$ to a few $\mathrm{kpc}$ (e.g., see Refs.~\cite{Jennings:2018psk,Mingarelli:2018myc,Antoniadis:2020gos} for recent pulsar distance determinations using \textit{Gaia} data). Moreover, current analysis methods treat the dark-matter-induced metric perturbation as a deterministic signal, imprinting a sinusoidal signature in the timing data of each pulsar.

In this work, we study the statistical properties of the dark-matter-induced metric perturbations to properly account for the spatial correlations of modes oscillating at a frequency $\sim 2m$~\cite{Khmelnitsky:2013lxt}. This builds on similar previous efforts in the literature~\cite{Luu:2023rgg} by calculating the two-point correlation function of the metric perturbations for galactic dark matter, which can be used to calculate the two-point correlation function of PTA timing residuals. Our statistical treatment provides a framework for improved PTA searches of ultralight dark matter.

\vspace{0.2cm}
\noindent \textbf{Impact of Dark Matter on Pulsar Timing.}
We consider metric perturbations induced by an ultralight scalar dark matter field.
In the weak-field limit, the metric $g_{\mu\nu}$ can be expanded around a Minkowski background through a small perturbation, $h_{\mu\nu}$:
\begin{equation}
  g_{\mu\nu} (t,\mathbf{x}) = \eta_{\mu\nu} + h_{\mu\nu} (t,\mathbf{x}) \, .
  \label{eq:g}
\end{equation}
We work with a mostly negative metric, such that $\eta_{\mu\nu} = \rm{diag} (+1, -1,-1,-1)$. Scalar metric perturbations can be written in Newtonian gauge in terms of two scalar potentials $\Phi$ and $\Psi$, such that
\begin{equation}
  h_{\mu\nu} = {\rm diag} (2\Phi , 2\Psi , 2\Psi , 2\Psi) \, .
  \label{eq:h}
\end{equation}
In Newtonian gauge, the effect on the timing of pulses observed from a pulsar due to scalar perturbations has the following three contributions:~\cite{Kim:2023kyy}
\begin{enumerate}
\item a difference in gravitational potential between the observer and the pulsar,
\item a difference in Doppler shifts due to accelerations experienced by the observer and pulsar, and
\item an integral of the perturbations along the photon worldline.
\end{enumerate}

The main observable for PTAs is the timing residual of light emitted by a pulsar. Both the Doppler and integral terms depend on the ratio of the spatial gradient of scalar perturbations to the oscillation frequency. In contrast, the potential term depends only on the scalar potential itself, not on its gradient. For oscillations of the dark matter field at frequencies $\sim 2m$, the Doppler and integral terms are suppressed due to the smallness of the dark matter velocity. Therefore, we focus primarily on the contribution from the potential term. 

Therefore, to leading order, the timing residual for pulsar $p$ at a location $\mathbf{x}_{p}$ is
\begin{equation}
  r_p (t) = \int^t \left[\Psi(t,\mathbf{0}) - \Psi(t - \left| \mathbf{x}_{p} \right|, \mathbf{x}_{p}) \right] dt \, ,
  \label{eq:ra}
\end{equation}
where we set the observer at the origin in our coordinate system and note that we omit the lower limit of integration, since it represents an arbitrary initial time. We take the Earth and pulsar to be at rest and discuss the impact of this assumption on our results in the Supplemental Material. It is worth emphasizing that although our main result for the statistical properties of Eq.~\eqref{eq:ra} is driven by the velocity dispersion of dark matter, the velocity independence of Eq.~\eqref{eq:ra} itself is what allows us to neglect the Doppler and integrated effects.

As shown in Ref.~\cite{Khmelnitsky:2013lxt}, for a non-relativistic plane wave of a scalar field with time-averaged density $\bar{\rho}$ and mass $m$, the potential is approximately
\begin{equation}
  \Psi \simeq \frac{\pi G \bar{\rho}}{m^2} \cos(2mt + \phi_0) \, ,
  \label{eq:psi}
\end{equation}
where $\phi_0$ is a uniformly distributed, random phase. Thus, in much of the literature on pulsar timing detection of dark matter, the signal is treated deterministically in time, with a sinusoidal time dependence for each pulsar~\cite{Porayko:2018sfa,EPTA:2023xxk,EPTA:2023xiy,NANOGrav:2023hvm}.

In this work, however, we are interested in the \textit{spatially stochastic} nature of the dark matter signal. Our main result is the evaluation of the two-point correlation of timing residuals
\begin{align}
  & \left\langle r_p(t) r_q(t') \right\rangle = \int^t dt \int^{t'} dt' ~ [\left\langle \Psi(t,\mathbf{0}) \Psi(t',\mathbf{0}) \right\rangle \label{eq:redcorr} \\
  & \quad + \left\langle \Psi(t - \left| \mathbf{x}_{p} \right|, \mathbf{x}_{p}) \Psi(t' - \left| \mathbf{x}_{q} \right|, \mathbf{x}_{q}) \right\rangle \notag \\
  & \quad - \left\langle \Psi(t - \left| \mathbf{x}_{p} \right|, \mathbf{x}_{p}) \Psi(t',\mathbf{0}) \right\rangle - \left\langle \Psi(t,\mathbf{0}) \Psi(t' - \left| \mathbf{x}_{q} \right|, \mathbf{x}_{q}) \right\rangle] \notag \, ,
\end{align}
which incorporates information about the spatial location of the observer and the pulsars $p$ and $q$. We show that this correlation function for timing residuals observed at $t$ and $t'$ is also deterministic in time, with a harmonic time dependence that behaves as $\cos[2m(t - t')]$. Therefore, while the main goal of this work is to study the spatial stochasticity of the PTA signal, it is still deterministic in time.

We can directly calculate the two-point correlation function of the timing residuals from the two-point correlation function of the metric perturbation, $\left\langle \Psi(t,\mathbf{x}) \Psi(t',\mathbf{x}') \right\rangle$. With a suitable choice of boundary conditions, the perturbation $\Psi$ can be derived using the 00 component of Einstein's equation with the weak-field expansion of Eqs.~\eqref{eq:g} and \eqref{eq:h},
\begin{equation}
  \nabla^2 \Psi = 4\pi G \rho \, ,
  \label{eq:poisson}
\end{equation}
where $G$ is Newton's constant and $\rho$ is the energy density of the scalar field, defined as the 00 component of the stress-energy tensor. Thus, the two-point correlation of the metric perturbation can be expressed in terms of the two-point correlation of the density field.

\vspace{0.2cm}
\noindent \textbf{Density Two-Point Function.}
For a scalar field $a(t,\mathbf{x})$ that obeys the Klein-Gordon equation, the stress-energy tensor is
\begin{equation}
  \tensor{T}{_\mu _\nu} = \partial_\mu a\, \partial_\nu a - \tensor{g}{_\mu _\nu} \left( \frac{1}{2} g^{\alpha\beta} \partial_\alpha a\; \partial_\beta a - \frac{1}{2} m^2 a^2 \right) \, .
\end{equation}
At zeroth order in the metric perturbations, the energy density of the field is
\begin{equation}
  \rho \equiv \tensor{T}{_0 _0} = \frac{1}{2} \left[ \dot{a}^2 + (\nabla a)^2 + m^2 a^2 \right] \, . \label{eq:Trho}
\end{equation}
Therefore, the density-density correlation function at two separated spacetime points $x \equiv (t,{\bf x})$ and $x' \equiv (t',{\bf x'}) $ is
\begin{align}
  \left\langle \rho(x) \rho(x') \right\rangle = \frac{1}{4} & \left\langle \left[ \dot{a}^2 + (\nabla a)^2 + m^2 a^2 \right] (x) \right. \\
  & \left. \left[ \dot{a}^2 + (\nabla a)^2 + m^2 a^2 \right] (x') \right\rangle \notag \, ,
\end{align}
where the angle brackets denote an ensemble average over the random variables that characterize the field. We provide details on the treatment of the field and its statistical properties in the Supplemental Material. The 9 four-point terms involving $a(x)$ and its derivatives in the above expression can each be expanded into 3 terms consisting of products of two-point functions, which are also given in the Supplemental Material. Assuming a time-independent and spatially homogeneous phase-space distribution function, $f(\mathbf{k})$, for the scalar field, we have
\begin{align}
  & \left\langle \rho(x) \rho(x') \right\rangle = \left\langle \rho(x) \rho(x') \right\rangle_0 + \int \frac{d^3k\, d^3k'}{(2\pi)^6 2 \omega_{\mathbf{k}} 2 \omega_{\mathbf{k}'}} f(\mathbf{k}) f(\mathbf{k}') \notag \\
  & \times \left\{A^+_{\mathbf{k} \mathbf{k}'} \cos \left[( \omega_{\mathbf{k}} + \omega_{\mathbf{k}'}) \Delta t - (\mathbf{k} + \mathbf{k}') \cdot \Delta \mathbf{x} \right] \right. \notag \\
  & \quad \left. + A^-_{\mathbf{k} \mathbf{k}'} \cos \left[( \omega_{\mathbf{k}} - \omega_{\mathbf{k}'}) \Delta t - (\mathbf{k} - \mathbf{k}') \cdot \Delta \mathbf{x} \right] \right\} \notag \, ,
  \label{eq:rhocorr2}
\end{align}
where $A_{\mathbf{k} \mathbf{k}'}^{\pm} \equiv (\omega_{\mathbf{k}} \omega_{\mathbf{k}'} + \mathbf{k} \cdot \mathbf{k}' \mp m^2)^2$, $\Delta t \equiv t-t'$, and $\Delta\mathbf{x} \equiv \mathbf{x}-\mathbf{x}'$. The quantity $\left\langle \rho(x) \rho(x') \right\rangle_0$ represents the collection of terms that are spacetime-separation independent, such as $\langle a(x)^2 \rangle \langle a(x')^2 \rangle$, and thus are not the focus of interest for this study.

The correlation depends on \textit{fast} modes (the frequency sum terms) and \textit{slow} modes (the frequency difference terms). We label these contributions to the correlation function with a subscript, e.g., $\left\langle \rho(x) \rho(x') \right\rangle_{\pm}$. While both modes are observable~\cite{Kim:2023pkx,Kim:2023kyy}, we focus on the fast modes, which have been the main target for PTA searches. Moreover, the expression for the timing residual in Eq.~\eqref{eq:ra} contains the leading contribution for the fast modes, but not the slow modes, which are dominated by the Doppler term~\cite{Kim:2023kyy}.

Working in the non-relativistic limit, the fast mode coefficient becomes $A_{\mathbf{k} \mathbf{k}'}^+ = m^4 \left(|\mathbf{v}+\mathbf{v}'|^2/2\right)^2$, up to corrections $\mathcal{O} (v^6)$. Additionally, we introduce the velocity phase space distribution $f(\mathbf{v})$, which is related to the momentum distribution through
\begin{equation}
  f(\mathbf{k}) \frac{d^3k}{(2\pi)^3} = \frac{\bar{\rho}}{m} f(\mathbf{v}) d^3v \, ,
\end{equation}
where we choose the normalization of $f(\mathbf{v})$ such that its integral over $d^3v$ is unity.
Thus, we can express the fast modes in the non-relativistic limit as
\begin{align}
  & \left\langle \rho(x) \rho(x') \right\rangle_+ = \frac{\bar{\rho}^2}{4} \int d^3v\, d^3v' f(\mathbf{v}) f(\mathbf{v}') \left(\frac{|\mathbf{v}+\mathbf{v}'|^2}{2}\right)^2 \notag \\
  &  \times \cos \left\{ m \left[ 2 + \frac{(v^2 + v^{\prime 2})}{2} \right] \Delta t - m (\mathbf{v} + \mathbf{v}') \cdot \Delta \mathbf{x} \right\} . \label{eq:rhocorr3}
\end{align}

To evaluate the integrals, we assume that the dark matter velocity is sampled from a Maxwell-Boltzmann distribution
\begin{equation}
  f(\mathbf{v}) \equiv \frac{1}{(2\pi v_0^2)^{3/2}} e^{-v^2 / 2 v_0^2}
  \label{eq:Maxwell-Boltzmann}
\end{equation}
as expected from the Standard Halo Model~\cite{Drukier:1986tm}. Since we work in the cosmic frame, the distribution function has a zero mean velocity. It is convenient to introduce a change of variables, $\mathbf{v}_\pm \equiv (\mathbf{v} \pm \mathbf{v}') / \sqrt{2}$, such that
\begin{align}
  & \left\langle \rho(x) \rho(x') \right\rangle_+ = \frac{\bar{\rho}^2}{4} \int d^3v_+ d^3v_- f(\mathbf{v}_-) f(\mathbf{v}_+) v_+^4 \label{eq:rhocorr4} \\
  & \quad \times \cos \left\{ m \left[2 + \frac{1}{2} (v_+^2 + v_-^2)\right] \Delta t - \sqrt{2} m \mathbf{v}_+ \cdot \Delta \mathbf{x} \right\} \, . \notag
\end{align}
The spatial dependence of the fast mode depends primarily on $\mathbf{v}_+$.

We carry out the individual integrals under a simplifying assumption: we neglect the $v_+^2 + v_-^2$ factor inside the cosine term of Eq.~\eqref{eq:rhocorr4}. This approximation is justified if we are interested in timescales short compared to the coherence time, $\ell / v_0 = (mv_0^2)^{-1}$. For ultralight dark matter with a mass around $10^{-23}~\mathrm{eV}$, the coherence time of the field exceeds a million years, making this an excellent approximation for typical PTA observations.
With this assumption, the fast mode correlation function becomes
\begin{align}
  \left\langle \rho(x) \rho(x') \right\rangle_+ & \simeq \frac{\bar{\rho}^2}{4m^4 \ell^4} \left(15 - 20 \frac{\left| \Delta \mathbf{x} \right|^2}{\ell^2} + 4 \frac{\left| \Delta \mathbf{x} \right|^4}{\ell^4} \right) \notag \\
  & \quad e^{-\left| \Delta \mathbf{x} \right|^2 / \ell^2} \cos\left(2 m \Delta t \right) \, , \label{eq:rhotwopoint}
\end{align}
where $\ell$ is the coherence length introduced in Eq.~\eqref{eq:coherence_length}. Note that the correlations are exponentially suppressed for separations larger than the coherence length.

The Fourier transform of the density is given by
\begin{equation}
  \rho(x) = \int \frac{d^3k}{(2\pi)^3} e^{i\mathbf{k} \cdot \mathbf{x}} \tilde{\rho} (t,\mathbf{k}) \, .
\end{equation}
Since the underlying scalar dark matter field is stationary and sampled from an isotropic distribution, the density two-point function takes the form
\begin{equation}
  \left\langle \tilde{\rho} (t,\mathbf{k}) \tilde{\rho } (t',\mathbf{k}') \right\rangle _+ = P_{+} (\Delta t ,k) (2\pi)^3 \delta^{(3)} (\mathbf{k} - \mathbf{k}') \, ,
\end{equation}
where $P_+(\Delta t,k)$ is the spatial power spectrum of the fast mode. From Eq.~\eqref{eq:rhotwopoint}, we find
\begin{equation}
  P_+(\Delta t, k) = \frac{\pi \sqrt{\pi}\bar{\rho}^2 k}{16 m^4} (k \ell)^3 e ^{-k^2 \ell^2 / 4} \cos(2m \Delta t) \, .
  \label{eq:Pplus}
\end{equation}

\vspace{0.2cm}
\noindent \textbf{Metric Perturbation Two-Point Function.}
The metric perturbation is related to the density through the Poisson equation, given in Eq.~\eqref{eq:poisson}. Using the Fourier transform of the density, the two-point correlation function for the gravitational potential $\Psi$ is
\begin{equation}
  \left\langle \Psi(x) \Psi(x') \right\rangle_+ = (4\pi G)^2 \int \frac{d^3k}{(2\pi)^3} \frac{P_{+}(\Delta t,k)}{k^4} e^{i\mathbf{k} \cdot \Delta \mathbf{x}} \, .
\end{equation}
The factor of $k^4$ in the denominator is a consequence of the long-range nature of gravitational interactions, which causes a divergence for density power spectra with significant support at low-momentum modes. This presents a conceptual issue for the slow modes, which can be addressed by incorporating a galactic density profile that decays with radial distance. However, for the fast mode, the Fourier transform is finite due to the factor of $k^4$ in front of the expression for $P_{+}(\Delta t,k)$ in Eq.~\eqref{eq:Pplus}.

Plugging in the density power spectrum and performing the integral yields the final result for the metric correlator:
\begin{equation}
  \left\langle \Psi(x) \Psi(x') \right\rangle_+ = \frac{\pi^2 G^2 \bar{\rho}^2}{m^4} e^{-|\Delta \mathbf{x}|^2/\ell^2} \cos(2m\Delta t) \, .
  \label{eq:psicorr}
\end{equation}
We observe that the spatial correlation of the metric perturbation is exponentially suppressed by the distance between the two different points in space.

The root mean square of the metric perturbation is
\begin{equation}
  \sqrt{\left\langle \Psi(x)^2 \right\rangle_+} = \frac{\pi G \bar{\rho}}{m^2} \, .
\end{equation}
This result is consistent with the variance obtained from Eq.~\eqref{eq:psi} and hence the result of Khmelnitsky and Rubakov~\cite{Khmelnitsky:2013lxt}.

\vspace{0.2cm}
\noindent \textbf{Timing Residual Two-Point Function.}
Substituting Eq.~\eqref{eq:psicorr} into Eq.~\eqref{eq:redcorr}, the two-point correlation of timing residuals is
\begin{align}
  & \left\langle r_p(t) r_q(t') \right\rangle_+ = \frac{\pi^2 G^2 \bar{\rho}^2}{4 m^6} \Big\{\cos(2m \Delta t)  \label{eq:redshiftcorr2} \\
  & -e^{-x_p^2/\ell^2} \cos\left[2m (\Delta t - x_p)\right] - e^{-x_q^2/\ell^2} \cos\left[2m (\Delta t + x_q) \right] \notag \\
  &  + e^{-|\mathbf{x}_p - \mathbf{x}_q|^2/\ell^2} \cos\left[2m (\Delta t - x_p + x_q)\right] \Big\} \notag \, ,
\end{align}
where $x_p \equiv \left| \mathbf{x}_p \right|$ and $x_q \equiv \left| \mathbf{x}_q \right|$. The terms in this expression represent the observer-observer correlation, the observer-pulsar $p$ correlation, the observer-pulsar $q$ correlation, and the pulsar $p$-pulsar $q$ correlation, respectively. The frequency of oscillations in this correlation function is $2m$, regardless of pulsar locations. This result is a consequence of neglecting the $v_+^2 + v_-^2$ in the argument of the cosine in Eq.~\eqref{eq:rhocorr4}.

In the $\ell \rightarrow 0$ limit, the exponential factors in the spatial correlation go to zero for any two separated points, leaving only the observer-observer term behind. The metric perturbations at different locations are uncorrelated, and the residual correlation function reduces to
\begin{equation}
\left\langle r_p(t) r_q(t') \right\rangle _+ = \frac{\pi^2 G^2 \bar{\rho}^2}{4m^6} \left(1 + \delta_{pq}\right) \cos(2m \Delta t) \, .
\end{equation}
For $p \neq q$, the only contribution comes from the observer-observer term, while for $p=q$, the correlation is enhanced by a factor of 2 due to the non-vanishing pulsar-pulsar term.

In the limit where $\ell \rightarrow \infty$, the exponential factors in the spatial correlation go to unity for all terms in the correlation function of residuals. Equation~\eqref{eq:redshiftcorr2} simplifies to
\begin{align}
  \left\langle r_p(t) r_q(t') \right\rangle _+ &= \frac{\pi^2 G^2 \bar{\rho}^2}{4m^6} \sin(mx_p) \sin(mx_q) \notag \\
  & \quad \cos\left[2m \Delta t + m (x_p - x_q) \right] \, ,
  \label{eq:redshiftcorr3}
\end{align}
which corresponds to the \textit{spatially deterministic} limit of the dark matter search. The covariance matrix derived from Eq.~\eqref{eq:redshiftcorr3} has two non-vanishing eigenvalues, regardless of the number of timing residuals for each pulsar. As such, the analysis requires marginalizing over only two independent random variables, which can be interpreted as the phase and amplitude of the dark matter field. Furthermore, if $x_p \rightarrow 0$ or $x_q \rightarrow 0$, the signal vanishes. This result is expected, since the pulsar and Earth terms in the correlation function of residuals cancel in this limit. 

From Eq.~\eqref{eq:redshiftcorr2}, it is clear that precise estimates of the pulsar locations are critical for conducting an optimal search for ultralight dark matter. The relative uncertainties in pulsar locations can vary significantly within a typical data set; some pulsar distances are known with percent-level precision, while others have uncertainties as large as the entire distance measurement. A practical approach to handle these uncertainties in a Bayesian analysis is to place a Gaussian prior on the pulsar distances. Our work does not attempt to quantify the sensitivity to pulsar locations, though it remains an important aspect for future investigations.

\vspace{0.2cm}
\noindent \textbf{Conclusions.}
The perturbations in the spacetime metric caused by dark matter present a unique opportunity for gravitational direct detection, potentially revealing the fundamental properties of dark matter, such as its mass and spin statistics~\cite{Antypas:2022asj}. Pulsar timing is currently the leading method for detecting these perturbations, with ongoing searches by the major collaborations focused on the mass range of $10^{-24} - 10^{-22}~\mathrm{eV}$. However, these searches rely on overly simplistic assumptions in the analysis, such as treating the metric perturbation at different pulsar locations as uncorrelated. These assumptions fall short, because the coherence length of dark matter is comparable to the distances between pulsars, meaning that pulsar timing signals exhibit correlations depending on the separations between pulsars and their distances to the observer.

In this work, we derive the two-point correlation function of timing residuals. In our framework, the effect of dark matter remains temporally deterministic but spatially stochastic. This spatial correlation function represents the dark matter analog of the Hellings-Downs angular correlation pattern of stochastic gravitational waves, and it is key in identifying whether an anomalous signal is due to dark matter or an unidentified source of noise.

These results are particularly timely, given recent results from the NANOGrav~\cite{NANOGrav:2023hvm} and EPTA~\cite{EPTA:2023xiy} collaborations, which identify unexplained monopolar signals around 4~nHz. If these signals were due to dark matter, it would correspond to a mass of approximately $8 \times 10^{-24}~\mathrm{eV}$. However, explaining the signals with dark matter would require a density near or exceeding the expected the local density estimate of $0.3~\mathrm{GeV/cm^3}$, which presents a challenge given the constraints on fuzzy dark matter from small-scale structure observations~\cite{Kobayashi:2017jcf, Irsic:2017yje, Nori:2018pka, Leong:2018opi, Schutz:2020jox, DES:2020fxi, Rogers:2020ltq,Dalal:2022rmp}. Regardless of the interpretation of these signals, our results suggest that existing treatments of the statistical properties of dark-matter-induced signals are inadequate, and we provide the groundwork for developing an analysis that would yield a more robust search of dark matter.

\begin{acknowledgments}
We thank Qiushi Wei for carefully reading the initial manuscript and providing useful comments. The research of JD is supported in part by the U.S. Department of Energy grant number DE-SC0025569. KB acknowledges support from the National Science Foundation under Grant No.~PHY-2413016.
\end{acknowledgments}

\appendix

\clearpage
\onecolumngrid
\newpage

\widetext
 \begin{center}
   \textbf{\large SUPPLEMENTAL MATERIAL \\[.2cm] ``On Pulsar Timing Detection of Ultralight Dark Matter''}\\[.2cm]
  \vspace{0.05in}
  {Kimberly K.~Boddy, Jeff A.~Dror, and Austin Lam}
\end{center}
\setcounter{equation}{0}
\setcounter{figure}{0}
\setcounter{table}{0}
\setcounter{page}{1}
\setcounter{section}{0}
\makeatletter
\renewcommand{\thesection}{S-\Roman{section}}
\renewcommand{\theequation}{S-\arabic{equation}}
\renewcommand{\thefigure}{S-\arabic{figure}}

\section{Scalar Field Two-Point Function}
To model the scalar field, we treat each constituent particle as a single plane wave labeled by an index $j$:
\begin{equation}
  a_j (t,\mathbf{x}) = a_0(\mathbf{k}) \cos(\omega_\mathbf{k} t - \mathbf{k} \cdot \mathbf{x} + \phi_{\mathbf{k},j}) \, ,
\end{equation}
where $\phi_{\mathbf{k},j} \in \left[0,2\pi\right)$, and the Klein-Gordon equation sets the dispersion relation $\omega^2_{\mathbf{k}} = \mathbf{k}^2 + m^2$. Each particle with momentum $\mathbf{k}$ has the same amplitude $a_0(\mathbf{k})$.

The statistical properties of the scalar field are determined from its phase-space distribution function, $f(\mathbf{k})$. We assume the distribution function is time-independent and spatially homogeneous. Note that the dark matter density in the Milky Way halo can vary over the kpc distances of typical pulsars in a PTA analysis, and we discuss this point further in the next section.

Consider particles in a small phase-space volume of size $d^3x\, d^3k$ around $\mathbf{x}$ and $\mathbf{k}$. The number of such particles is $N(\mathbf{k}) = f(\mathbf{k})\, d^3x\, d^3k / (2\pi)^3$, and these particles make up the field
\begin{equation}
  a_{\mathbf{k}}(t,\mathbf{x}) = a_0(\mathbf{k}) \sum_{j=1}^{N(\mathbf{k})} \cos(\omega_{\mathbf{k}}t - \mathbf{k} \cdot \mathbf{x} + \phi_{\mathbf{k},j}) \, .
\end{equation}
Summing over uniformly distributed phases $\phi_{\mathbf{k},j}$ gives a new uniformly distributed phase $\phi_{\mathbf{k}}$ and a Rayleigh-distributed variable $\alpha_{\mathbf{k}}$ with scale parameter $1$~\cite{Foster:2017hbq,Centers:2019dyn,Dror:2021nyr}:
\begin{equation}
  \sum_{j=1}^{N(\mathbf{k})} e^{i\phi_{\mathbf{k},j}} = \alpha_{\mathbf{k}} \sqrt{\frac{N(\mathbf{k})}{2}} e^{i\phi_{\mathbf{k}}} \, .
\end{equation}
Thus,
\begin{equation}
  a_{\mathbf{k}}(t,\mathbf{x}) = \sqrt{\frac{f(\mathbf{k})\, d^3k\, d^3x}{2(2\pi)^3}} a_0(\mathbf{k}) \alpha_{\mathbf{k}} \cos(\omega_{\mathbf{k}}t - \mathbf{k}\cdot \mathbf{x} + \phi_{\mathbf{k}}) \, .
\end{equation}
Since we are interested in the statistical properties of this field, we introduce angle brackets $\langle \rangle$ to denote an ensemble average, obtained by integrating over the random variables $\phi_{\mathbf{k}}$ and $\alpha_{\mathbf{k}}$, weighted by their respective distribution functions.

We fix the amplitude $a_0(\mathbf{k})$ by considering the energy density of this collection of plane waves:
\begin{equation}
  \rho_{\mathbf{k}} = \frac{f(\mathbf{k})\, d^3k\, d^3x}{2(2\pi)^3} a_0(\mathbf{k})^2 \alpha_{\mathbf{k}}^2 \left\{ \omega_{\mathbf{k}}^2 - \frac{k^2}{2} \cos\left[2(\omega_{\mathbf{k}}t - \mathbf{k} \cdot \mathbf{x} + \phi_\mathbf{k}) \right] \right\} \, .
\end{equation}
We set the ensemble average of this quantity to the thermodynamic energy density of $N(\mathbf{k})$ particles, $\langle \rho_{\mathbf{k}} \rangle = \omega_{\mathbf{k}} f(\mathbf{k}) d^3k / (2\pi)^3$, giving the normalization condition,
\begin{equation}
  a_0(\mathbf{k}) = \sqrt{\frac{2}{d^3x\, \omega_{\mathbf{k}}}} \, .
\end{equation}
Therefore,
\begin{equation}
  a_{\mathbf{k}}(t,\mathbf{x}) = \sqrt{\frac{f(\mathbf{k})\, d^3k}{(2\pi)^3 \omega_{\mathbf{k}}}} \alpha_{\mathbf{k}} \cos(\omega_{\mathbf{k}} t - \mathbf{k} \cdot \mathbf{x} + \phi_{\mathbf{k}}) \, .
\end{equation}
Summing over all momenta, the total field is
\begin{equation}
  a(t,\mathbf{x}) = \sum_{\mathbf{k}} \sqrt{\frac{f(\mathbf{k}) d^3k}{(2\pi)^3 \omega_{\mathbf{k}}}} \alpha_{\mathbf{k}} \cos(\omega_{\mathbf{k}}t - \mathbf{k} \cdot \mathbf{x} + \phi_{\mathbf{k}}) \, .
  \label{eq:axionfield}
\end{equation}
Since the only remaining random variables are $\phi_{\mathbf{k}}$ and $\alpha_{{\mathbf{k}}}$, the scalar field is a Gaussian random field.

From Eq.~\eqref{eq:axionfield}, the two-point correlation function of the field at two spacetime points is
\begin{equation}
  \left\langle a(x) a(x') \right\rangle = \sum_{\mathbf{k},\mathbf{k}'} \sqrt{\frac{f(\mathbf{k}) f(\mathbf{k}') d^3k\, d^3k'}{(2\pi)^6 \omega_{\mathbf{k}} \omega_{\mathbf{k}'}}} \left\langle \alpha_{\mathbf{k}} \alpha_{\mathbf{k}'} \cos(\omega_{\mathbf{k}}t - \mathbf{k} \cdot \mathbf{x} + \phi_{\mathbf{k}}) \cos(\omega_{\mathbf{k}'}t' - \mathbf{k}' \cdot \mathbf{x}' + \phi_{\mathbf{k}'})\right\rangle \, ,
\end{equation}
where we denote $a(x) \equiv a(t,\mathbf{x})$. If $\mathbf{k} \neq \mathbf{k}'$, the expectation value in the sum is zero since $\left\langle \cos(\omega_{\mathbf{k}}t - \mathbf{k} \cdot \mathbf{x} + \phi_{\mathbf{k}}) \right\rangle = 0$. Otherwise, it is given by a cosine of the difference in the arguments of the cosines. The final result is,
\begin{align}
  \left\langle a(x) a(x') \right\rangle & \rightarrow \int \frac{d^3k f(\mathbf{k})}{(2\pi)^3 \omega_{\mathbf{k}}} \cos(\omega_{\mathbf{k}} \Delta t - \mathbf{k} \cdot \Delta \mathbf{x}) \, ,
\end{align}
where we have promoted the sum to an integral and introduced the notation $\Delta t \equiv t - t'$ and $\Delta \mathbf{x} \equiv \mathbf{x} - \mathbf{x}'$. In deriving this expression, we assume the phase space density is independent of time and space.

We similarly calculate other relevant two-point functions to be
\begin{align}
  \left\langle \dot{a}(x) a(x') \right\rangle &= -\int \frac{d^3k}{(2\pi)^3 \omega_{\mathbf{k}}} \omega_{\mathbf{k}} f(\mathbf{k}) \sin(\omega_{ \mathbf{k}}(t - t') - \mathbf{k} \cdot (\mathbf{x} - \mathbf{x}')) \\
  \left\langle \nabla a(x) a(x') \right\rangle &= \int \frac{d^3k}{(2\pi)^3 \omega_{\mathbf{k}}} \mathbf{k} f(\mathbf{k}) \sin(\omega_{\mathbf{k}}(t - t') - \mathbf{k} \cdot (\mathbf{x} - \mathbf{x}')) \\
  \left\langle \dot{a}(x) \dot{a}(x') \right\rangle &= \int \frac{d^3k}{(2\pi)^3 \omega_{\mathbf{k}}} \omega^2_{\mathbf{k}}  f(\mathbf{k}) \cos(\omega_{\mathbf{k}}(t - t') - \mathbf{k} \cdot (\mathbf{x} - \mathbf{x}')) \\
  \left\langle \partial_i a(x) \partial_j a(x') \right\rangle &= \int \frac{d^3k}{(2\pi)^3 \omega_{\mathbf{k}}} k_i k_j f(\mathbf{k}) \cos(\omega_{\mathbf{k}}(t - t') - \mathbf{k} \cdot (\mathbf{x} - \mathbf{x}')) \\
  \left\langle \dot{a}(x) \nabla a(x') \right\rangle &= -\int \frac{d^3k}{(2\pi)^3 \omega_{\mathbf{k}}} \omega_{\mathbf{k}}\mathbf{k} f(\mathbf{k}) \cos(\omega_{\mathbf{k}}(t - t') - \mathbf{k} \cdot (\mathbf{x} - \mathbf{x}'))
\end{align}
Furthermore, since $a(x)$ is a Gaussian random field, we can reduce higher-point functions into products of two-point functions, such as
\begin{equation}
  \left\langle a(x_1) a(x_2) a(x_3) a(x_4) \right\rangle = \left\langle a(x_1) a(x_2) \right\rangle \left\langle a(x_3) a(x_4) \right\rangle + \left\langle a(x_1) a(x_3) \right\rangle \left\langle a(x_2) a(x_4) \right\rangle + \left\langle a(x_1) a(x_4) \right\rangle \left\langle a(x_2) a(x_3) \right\rangle \label{eq:4to2} \, .
\end{equation}

\section{Discussion of Assumptions}
\subsection{Fixed Earth and Pulsar Locations}
In the expression for the timing residual in Eq.~\eqref{eq:ra}, we treat the Earth and pulsar as if they are at rest. Relaxing this approximation would modify the metric perturbation evaluated at the pulsar to be
\begin{equation}
  \Psi(t - \left| \mathbf{x}_{p} \right|, \mathbf{x}_{p}) \rightarrow \Psi( (1- \mathbf {v} \cdot \hat {\bf x}_{p} )t-|\mathbf {x} _{p}|, \mathbf{x}_p +\mathbf{v}t) \, ,
\end{equation}
where $\mathbf{x}_p$ is interpreted as the position at a reference time. This modification would generate corrections $\mathcal{O} (m v x_p)$ in the cosine arguments of Eq.~\eqref{eq:redshiftcorr2}. Since $v \sim v_0$, the corrections are $\mathcal{O} (1)$ when the correlation length is of order the pulsar distances. In practice, the $x_p$ values are not known to sufficient precision to make this effect necessary to include. Instead, one may simply take the cosines of the observer-pulsar $p$, observer-pulsar $q$, and pulsar $p$-pulsar $q$ correlations to all have a randomly sampled phase.

\subsection{Uniform Dark Matter Density}
For a Navarro-Frenk-White profile with a Maxwell-Boltzmann velocity distribution given in Eq.~\eqref{eq:Maxwell-Boltzmann}, the density variations may be significant. Within a kpc, the background density varies by about 20\%, while the velocity dispersion $v_0$ varies by about a 2\% percent. As such, the variation in density can be approximately incorporated into Eq.~\eqref{eq:redshiftcorr2} by replacing the overall $\bar{\rho}^2$ with the product of the time-averaged densities at the two locations contributing to each term in the residual correlation function:
\begin{align}
  & \left\langle r_p(t) r_q(t') \right\rangle_+ = \frac{\pi^2 G^2}{4m^6} \bigg[ \bar{\rho} (\mathbf{0})^2 \cos(2m \Delta t) \\
  & \quad - \bar{\rho}(\mathbf{0}) \bar{\rho}(\mathbf{x}_p) e^{-x_p^2/\ell^2} \cos\left[2m(\Delta t - x_p)\right] - \bar{\rho}(\mathbf{0}) \bar{\rho}(\mathbf{x}_q) e^{-x_q^2/\ell^2} \cos\left[2m(\Delta t + x_q)\right] \notag \\
  & \quad + \bar{\rho}(\mathbf{x}_p) \bar{\rho}(\mathbf{x}_q) e^{-|\mathbf{x}_p - \mathbf{x}_q|^2/\ell^2} \cos\left[2m(\Delta t - x_p + x_q)\right] \bigg] \notag \, .
\end{align}

\bibliography{references}

\end{document}